\documentclass{iau}
\usepackage{graphicx}
\usepackage{epsfig} 

\def\farcs{\hbox{$.\!\!^{\prime\prime}$}}

\title[High-resolution SNR1987A at high radio frequencies] 
{The radio remnant of Supernova 1987A \\at high frequencies and high resolution}

\author[Zanardo et al.]
{G. Zanardo$^1$, L. Staveley-Smith$^{1,4}$, C. -Y. Ng$^2$, B. M. Gaensler$^{3,4}$\\
T. M. Potter$^1$, R. N. Manchester$^5$ \and A. K. Tzioumis$^5$}

\affiliation{
$^1$International Centre for Radio Astronomy Research (ICRAR), \\ M468, The University of Western Australia, Crawley, WA 6009, Australia. \\ email: {\tt giovanna.zanardo@icrar.org} \\[\affilskip]
$^2$Department of Physics, The University of Hong Kong, \\ Pokfulam Road, Hong Kong \\[\affilskip]
$^3$Sydney Institute for Astronomy (SIfA), School of Physics, The University of Sydney, \\NSW 2006, Australia \\[\affilskip]
$^4$Australian Research Council Centre of Excellence for All-sky Astrophysics (CAASTRO) \\[\affilskip]
$^5$CSIRO Astronomy and Space Science, Australia Telescope National Facility, \\PO Box 76, Epping, NSW 1710, Australia}

\pubyear{2013}
\volume{296}  
\pagerange{119--126}
\jname{Supernova Environmental Impacts}
\editors{A. K. Ray \& R. McCray, eds.}
\begin{document}

\maketitle

\begin{abstract}
As the remnant of Supernova (SN) 1987A has been getting brighter over time, new observations at high frequencies have allowed imaging of the radio emission at unprecedented detail. We present a new radio image at 44 GHz of the supernova remnant (SNR), derived from observations performed with the Australia Telescope Compact Array (ATCA) in 2011. The diffraction-limited image has a resolution of $349\times225$ mas, which is the highest achieved to date in high-dynamic range images of the SNR. We also present a new image at 18 GHz, also derived from ATCA observations performed in 2011, which is super-resolved to $0\farcs25$. The new 44 and 18 GHz images yield the first high-resolution spectral index map of the remnant. The comparison of the 44 GHz image with contemporaneous X-ray and H$\alpha$ observations allows further investigations of the nature of the remnant asymmetry and sheds more light into the progenitor hypotheses and SN explosion. In light of simple free-free absorption models, we discuss the likelihood of detecting at 44 GHz the possible emission originating from a pulsar wind nebula (PWN) or a compact source in the centre of the remnant.
\end{abstract}

\firstsection 
\vspace{-0.0mm}              
\section{Introduction}
Supernova 1987A in the Large Magellanic Cloud, as the only nearby  core-collapse  supernova observed to date, has provided a unique opportunity to study the evolution of the interaction between the propagating blast wave and the progenitor's circumstellar medium (CSM). The complex CSM distribution in the supernova remnant (SNR) is believed to have originated from a red supergiant (RSG), which has evolved into a blue supergiant (BSG) about 20,000 years before the explosion (\cite[Crotts \& Heathcote 2000]{cro00}). Models of the progenitor evolution suggest that the equatorially denser, i.e.\ slower, RSG wind (\cite[Blondin \& Lundqvist 1993]{blo93}, \cite[Martin \& Arnett 1995]{mar95}), was swept up by the faster BSG wind (\cite[Morris \& Podsiadlowski 2007]{mor07}), thus forming high density rings. In particular, beside the central circular ring in the equatorial plane ({\it equatorial ring}, ER), observations have also revealed two outer rings that formed from the mass loss of the progenitor star,  located on either side of the equatorial plane (\cite[Jakobsen et al. 1991]{jak91}, \cite[Plait et al. 1995]{pla95}), which confer to SNR 1987A a peculiar triple-ring nebula structure. \\
Since the radio detection of the remnant in mid-1990  (\cite[Turtle et al. 1990]{tur90}), the synchrotron emission has been generated by the shock wave propagating into the ring-shaped distribution of the CSM in the equatorial plane. Monitoring of the flux density has been regularly undertaken with the Molonglo Observatory Synthesis Telescope (MOST) at 843 MHz  (\cite[Ball et al. 2001]{bal01}) and at 1.4, 2.4, 4.8 and 8.6 GHz with the Australia Telescope Compact Array (ATCA) (\cite[Staveley-Smith et al. 1992]{sta92}). ATCA observations have been ongoing for $\sim$25 years (\cite[Staveley-Smith et al.1993]{sta93}, \cite[Gaensler et al. 1997]{gae97}, \cite[Manchester et al. 2002]{man02}, \cite[Staveley-Smith et al.  2007]{sta07}, \cite[Zanardo et al. 2010]{zan10}). An exponential increase of the flux density has been measured at all frequencies since day $\sim$5000 after the explosion, which is likely due to an increasing efficiency of the acceleration process of particles by the shock front (\cite[Zanardo et al. 2010]{zan10}).
The morphology of the nonthermal radiation emitted by relativistic electrons accelerated in the remnant has been investigated using images at 9 GHz since 1992 (\cite[Staveley-Smith et al. 1992]{sta92}), with a spatial resolution of 0\farcs5  achieved via maximum entropy super-resolution (\cite[Gaensler et al. 1997]{gae97}, \cite[Ng et al. 2008]{ng08}).  These images have provided the first insight into the marked east-west asymmetry of the radio emission. 
The first imaging observations at 18 GHz were undertaken in 2003 July, at an effective resolution of 0\farcs45 (\cite[Manchester et al. 2005]{man05}). Very long baseline interferometry (VLBI) observations of the SNR were successful in 2007 October (\cite[Tingay et al. 2009]{tin09}) and 2008 November  (\cite[Ng et al. 2011]{ng11}) at 1.4 and 1.7 GHz, respectively. These observations, while with low sensitivity and dynamic range, at a resolution of $\sim0\farcs1$ captured the presence of structures smaller than $0\farcs2$ in bright regions (\cite[Ng et al. 2011]{ng11}). At the same time, ATCA observations at 36 GHz in 2008 April and October resulted in high-dynamic range images with an angular resolution of 0\farcs3 (\cite[Potter et al. 2009]{pot09}).  
After the ATCA upgrade in mid-2009 with the Compact Array Broadband Backend (\cite[Wilson et al. 2011]{wil11}), the remnant was  imaged at higher frequencies. The first resolved image at 94 GHz was produced from observations between 2011 June and August  by \cite[Laki\'cevi\'c et al. (2012)]{lak12}. 

\vspace{-4mm}
\section{New observations}
\label{new}
\begin{figure*}[htb]
\begin{minipage}[c]{62.3mm}
\begin{center}
\vspace{-0.1mm}
\advance\leftskip75.0mm
\includegraphics[width=62.3mm, angle=0]{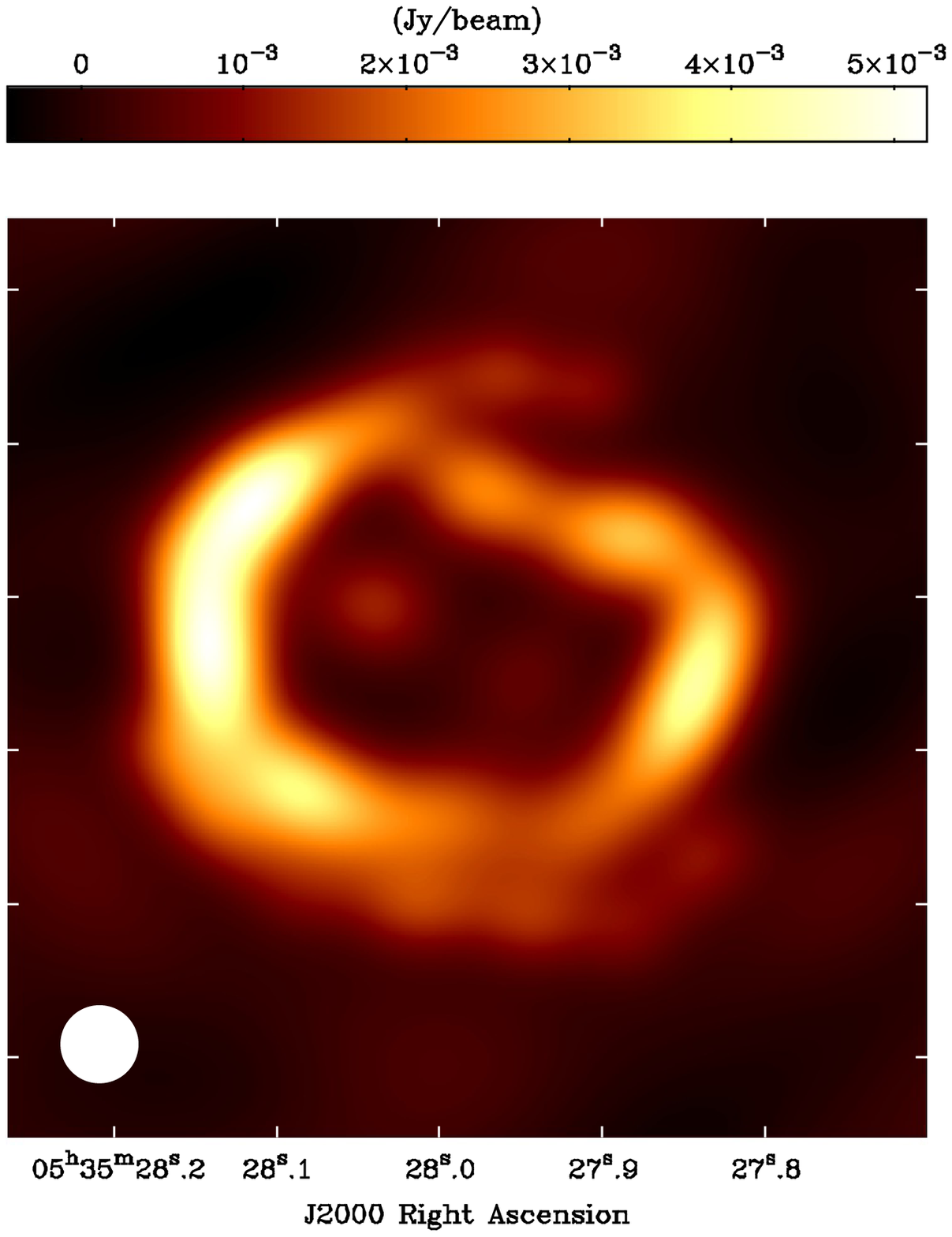}
\end{center}
\end{minipage}
\begin{minipage}[c]{77.7mm}
\begin{center}
\vspace{0.5mm}
\advance\leftskip-135mm
\includegraphics[width=77.7mm, angle=0]{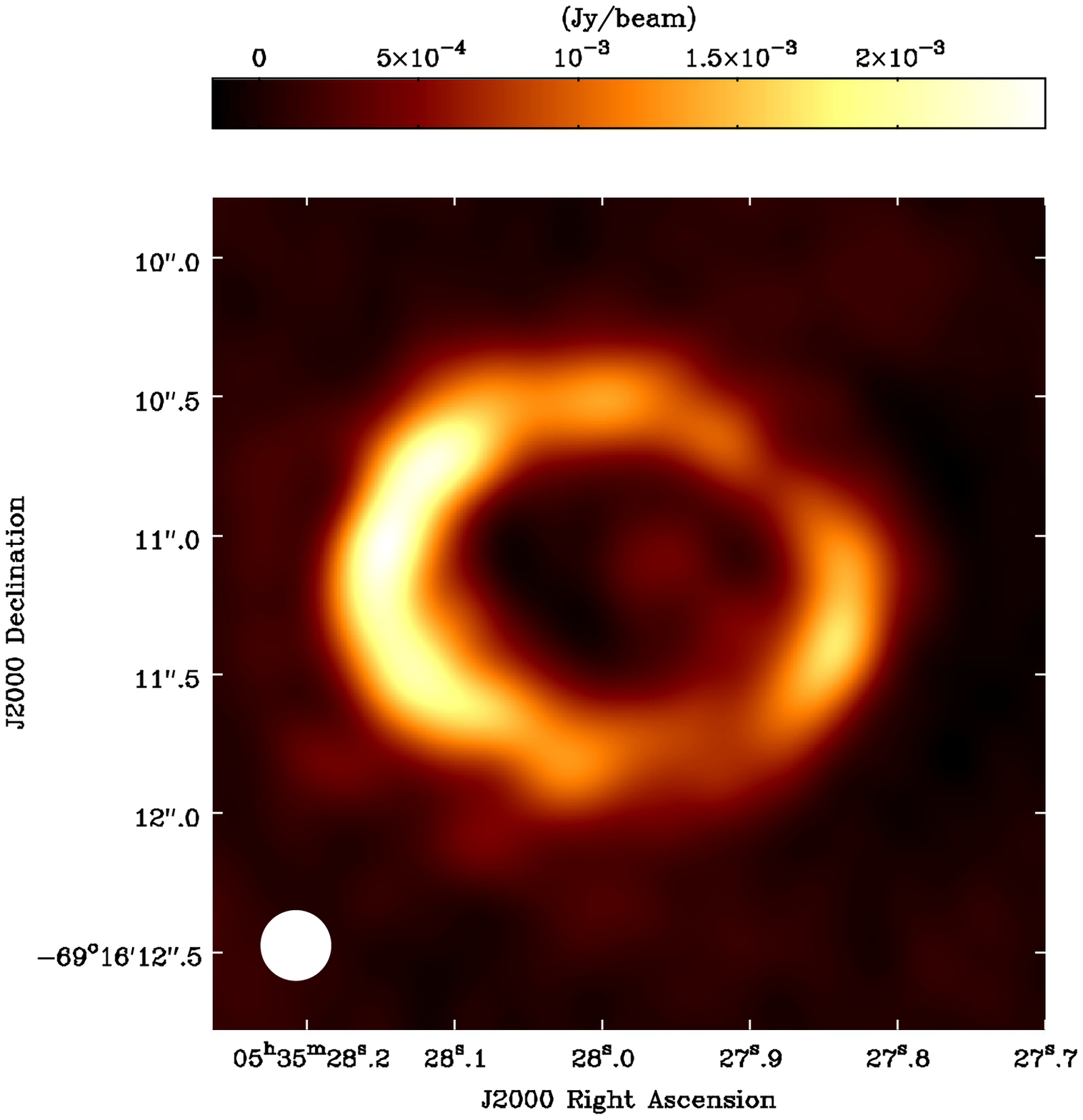}
\end{center}
\end{minipage}
\caption{{\it Left:} Super-resolved Stokes-I continuum image of SNR 1987A at 44 GHz made by combining observations performed with the ATCA on 2011 January 24 and November 16. The diffraction-limited image is restored with a $0\farcs25$ circular beam. {\it Right:} Super-resolved Stokes-I continuum 18 GHz image obtained from observations performed in 2011 January using a $0\farcs25$ restoring circular beam. For details on both images see \cite[Zanardo et al. (2013)]{zan13}.}
\label{fig1}
\vspace{-0mm}
\end{figure*}

The first image of the SNR at 44 GHz was derived from ATCA observations in 2011 January and November, with diffraction-limited resolution of $349\times225$ mas, which is the highest achieved to date at high-dynamic range (\cite[Zanardo et al. 2013]{zan13}). This new image has been analysed in conjunction with that derived from contemporaneous ATCA observations at 18 GHz (\cite[Zanardo et al. 2013]{zan13}) to investigate the spectral index distribution across the remnant. In Fig.\,\ref{fig1}, the 44 GHz image, slightly super-resolved to $0\farcs25$, is shown together with the super-resolved 18 GHz image. \\
The high resolution of the 44 GHz image has allowed  a comparison between the morphology of the radio and H$\alpha$ emission, as seen in   contemporaneous observations with  the  Hubble Space Telescope ({\it HST})\footnote[1]{STScI-2011-21, NASA, ESA, \& Challis P. (Harvard--Smithsonian Center for Astrophysics)
http://hubblesite.org/newscenter/archive/releases/2011/21/image/},
in terms of both emission around the ER and emission from the centre of the SNR, where the ejecta can be located (\cite[Larsson et al. 2011]{lar11}). 
\vspace{-0.mm}
%
%
Comparison with the X-ray observations performed with the {\it Chandra} Observatory (\cite[Helder et al. 2013]{hel13}) has provided further insight into the structure of the nonthermal emission.

\vspace{-4.mm}
\section{Findings}
\label{Over}

{\underline{\it Emission morphology and asymmetry}}. Consistent with previous radio observations, the 44 GHz image shows a marked asymmetry in the emission distribution.  More specifically, the east-west asymmetry ratio is $\sim1.5$ from the ratio of the brightness peaks in the radial profiles at PA $\sim$90$^{\circ}$, is $\sim1.6$ from the integrated flux densities over the eastern and western halves of the image.
These values are higher than the $\sim$1.4 ratio derived for the new 18 GHz image and the ratio previously measured with images at lower frequencies.

{\underline{\it Remnant expansion and local velocities}}. The comparison between the new images at both 18 and 44 GHz with corresponding observations performed in earlier epochs, specifically the  2003 observations at 17 and 19 GHz and the 2008 observations at 36 GHz, highlights an asymmetric expansion of the remnant, with expansion velocities on the eastern lobe significantly higher than what measured on the western lobe.

{\underline{\it Spectral index variations}}. The 18--44 GHz spectral index distribution is measured at an angular resolution of $0\farcs4$. The spectral indices in SNR 1987A primarily range between $-1.1$ and $-0.3$, with a mean of $-0.8$. Spectral indices associated with the brightest sites over the eastern lobe are steeper than the mean value.  The steeper spectrum on the eastern lobe implies  compression ratios slightly lower than on the western bright sites,  and could be correlated with the higher expansion rate measured on the eastern side of the remnant. Two regions of flatter spectral indices are identified, one approximately located in the centre of the SNR, which extends over the SN location (\cite[Reynolds et al. 1995]{rey95}), and the other located further north. These two features lie at PA $\sim30^{\circ}$.

{\underline{\it Structure of the shock and progenitor explosion}}. There is a strong correspondence between major features of the emission at 44 GHz, and the arrangement of the hot spots shown in the H$\alpha$ emission. The direction of the east-west asymmetry of the X-ray and radio emission, is opposite to that of the H$\alpha$ emission. This fact supports the hypothesis that the remnant asymmetric morphology might be due to an asymmetric explosion, rather than to an asymmetric distribution of the CSM.

{\underline{\it Likelihood of detecting the radiation emitted by a compact source}}. At 44 GHz, a central feature of fainter emission appears to extend over the SN site, and to overlap with the western side of the ejecta as seen by {\it HST}. This feature corresponds to a region of flatter spectral indices in the 18--44 GHz spectral map, which could indicate the presence of a compact source or a PWN. The origin of this emission is unclear.
However, simple free-free absorption models suggest that the radiation emitted by a compact source inside the equatorial ring may now be detectable at high frequencies, or at lower frequencies if there are holes in the ionised component of the ejecta. 
Future high-resolution observations, both at lower frequencies with VLBI and at higher frequencies with ATCA and the Atacama Large sub-Millimeter Array,  will be crucial to further investigate the nature of this emission.






\vspace{-3.mm}
\begin{thebibliography}{}

\bibitem[Ball et al.(2001)]{bal01} Ball, L., Crawford, D.F., Hunstead, R. W., Klamer, I., \& McIntyre, V. J. 2001, \textit{ApJ}, 549, 599
\bibitem[Blondin \& Lundqvist, (1993)]{blo93} Blondin, J.~M., Lundqvist, P. 1993, \textit{ApJ},405, 337
\bibitem[Crotts \& Heathcote(2000)]{cro00} Crotts, A. P. S., \& Heathcote, R. S. 2000, \textit{ApJ}, 528, 426
\bibitem[Gaensler et al.(1997)]{gae97}  Gaensler, B. M., Manchester, R. N., Staveley-Smith, L., Tzioumis, A. K., Reynolds, J. E., \& Kesteven, M. J. 1997,  \textit{ApJ}, 479, 845
\bibitem[Gaensler(1998)]{gae98} Gaensler, B.~M. 1998, \textit{ApJ}, 493, 781
\bibitem[Gaensler et al.(2007)] {gae07} Gaensler, B. M., Staveley-Smith, L., Manchester, R. N., Kesteven, M. J., Ball, L., \& Tzioumis, A. K. 2007,  in AIP Conf. Proc. 937, Supernova 1987A: 20 Years After: Supernovae and Gamma-Ray Bursters, ed. S. Immler, K. W. Weiler, \& R. McCray (New York: AIP), 86 
\bibitem[Helder et al.(2013)]{hel13} Helder, E. A., Broos, P. S., Dewey, D., Dwek, E., McCray, R., Park, S., Racusin, J. L., Zhekov, S. A., \& Burrows, D. N. Ê2013, \textit{ApJ}, 764,Ê11
\bibitem[Jakobsen et al.(1991)]{jak91} Jakobsen, P. et al. 1991, \textit{ApJ}, 369, 63
\bibitem[Laki\'cevi\'c et al.(2012)]{lak12} Laki\'cevi\'c, M., Zanardo, G., van Loon, J.~Th., Staveley-Smith, L., Potter, T., Ng, C.~-Y., \& Gaensler, B.~M.  2012, \textit{A\&A}, 541, L2
\bibitem[Larsson et al.(2011)]{lar11} Larsson, J. et al. 2011, \textit{Nature}, 474, L484
\bibitem[Manchester et al.(2002)]{man02} Manchester, R. N., Gaensler, B. M., Wheaton, V. C., Staveley-Smith,  L., Tzioumis, A. K., Bizunok, N. S., Kesteven, M. J., \& Reynolds, J. E. 2002, \textit{PASA}, 19, 207 
\bibitem[Manchester et al.(2005)]{man05} Manchester, R.~N., Gaensler, B.~M., Staveley-Smith, L.,  Kesteven, M.~J. \& Tzioumis, A.~K. 2005, \textit{ApJ}, 628, L131
\bibitem[Martin \& Arnett(1995)]{mar95} Martin, C.~L., \& Arnett, D. 1995 \textit{ApJ}, 447, 378
\bibitem[Morris \& Podsiadlowski(2007)]{mor07} Morris, T., \& Podsiadlowski, P. 2007, \textit{Science}, 315, 1103 
\bibitem[Ng et al.(2008)]{ng08} Ng, C.-Y., Gaensler, B. M., Staveley-Smith, L., Manchester, R. N., Kesteven,  M. J., Ball, L., \& Tzioumis, A. K. 2008, \textit{ApJ}, 684, 481
\bibitem[Ng et al.(2011)]{ng11} Ng, C. -Y. et al. 2011, \textit{ApJ} (Letters), 728, L15
\bibitem[Plait et al.(1995)]{pla95} Plait, P.~C., Lundqvist, P., Chevalier, R.~A., \& Kirshner, R.~P. 1995, \textit{ApJ}, 439, 730
\bibitem[Podsiadlowski, Morris \& Ivanova(2007)]{pod07} Podsiadlowski, Ph., Morris, T.~S., \& Ivanova, N. 2007, AIP Conf. Proc., 937, 125-133
\bibitem[Potter et al.(2009)]{pot09} Potter, T. M., Staveley-Smith, L., Ng, C.-Y., Ball, Lewis, Gaensler, B. M., Kesteven, M. J., Manchester, R. N., Tzioumis, A. K., \&  Zanardo, G. 2009, \textit{ApJ}, 705, 261
\bibitem[Reynolds et al.(1995)]{rey95} Reynolds, J.~E. et al. 1995, \textit{A\&A}, 304, 116
\bibitem[Staveley-Smith et al.(1992)]{sta92} Staveley-Smith, L. et al. 1992, \textit{Nature}, 355, 147
\bibitem[Staveley-Smith et al.(1993)]{sta93}Staveley-Smith, L. et al. 1993, \textit{Nature}, 366, 136
\bibitem[Staveley-Smith et al.(2007)]{sta07} Staveley-Smith, L., Gaensler, B.~M., Manchester, R.~N., Ball, L., Kesteven, M.~J., \& Tzioumis, A.~K.\ 2007, in AIP Conf. Proc. 937, Supernova 1987A: 20 Years After: Supernovae and Gamma-Ray Bursters, ed. S. Immler, K.W. Weiler, \& R. McCray, (New York: AIP), 96 
\bibitem[Tingay et al.(2009)]{tin09} Tingay, S. et al. 2009, in $8^{\rm th}$ International e-VLBI Workshop, Proc. Sci., 100.
\bibitem[Turtle et al.(1990)]{tur90} Turtle, A. J., Campbell-Wilson, D., Manchester, R.N., Staveley-Smith, L., \& Kesteven, M. J. 1990, IAU Circ. 5086, 2
\bibitem[Wilson W.E. et al. (2011)]{wil11} Wilson W.~E. et al., \textit{MNRAS}, 416, 832
\bibitem[Zanardo et al.(2010)]{zan10} Zanardo, G., Staveley-Smith, L., Ball, Lewis, Gaensler, B. M., Kesteven, M. J., Manchester, R. N., Ng, C.-Y., Tzioumis, A. K., \& Potter, T. M. 2010, \textit{ApJ}, 710, 1515
\bibitem[Zanardo et al.(2013)]{zan13} Zanardo, G., Staveley-Smith, L., Ng, C.-Y., Gaensler, B. M., Potter, T. M. Manchester, R. N., \& Tzioumis, A. K. 2013,   \textit{ApJ}, 767, 98

\end{thebibliography}
\end{document}